\documentclass[utf8]{frontiersSCNS} 

\usepackage{url,hyperref,lineno,microtype,subcaption}
\usepackage[onehalfspacing]{setspace}

\newcommand{\nt}{}

\def\keyFont{\fontsize{8}{11}\helveticabold }
\def\firstAuthorLast{Nokhrina} 
\def\Authors{Nokhrina Elena\,$^{*}$}
\def\Address{Relativistic Astrophysics Laboratory: Basic and Applied Studies of Space Objects, Moscow Institute of Physics and Technology, School of Fundamental and Applied Physics, Dolgoprudny, Russia}
\def\corrAuthor{Institutsky per. 9, Dolgoprudny, 141700, Russia}

\def\corrEmail{nokhrina@phystech.edu}

\begin{document}

\onecolumn
\firstpage{1}

\title[Jet magnetic flux]{The correlation between the total magnetic flux and the total jet power} 

\author[\firstAuthorLast ]{\Authors} 
\address{} 
\correspondance{} 

\extraAuth{}

\maketitle

\begin{abstract}

\section{}
Magnetic field threading a black hole ergosphere is believed to play the key role in both
driving the powerful relativistic jets observed in active galactic nuclei and extracting the 
rotational energy from a black hole via Blandford-Znajek process.
The magnitude of magnetic field and the magnetic flux in the vicinity of a central black hole is 
predicted by theoretical models.
On the other hand, the magnetic field in a jet
can be estimated through measurements of either the core shift effect or the brightness temperature.
In both cases the obtained magnetic field is in the radiating domain, so its direct application to the calculation 
of the magnetic flux needs some theoretical assumptions.
In this paper we address the issue of estimating the magnetic flux contained in a jet using the measurements of 
{a core shift effect and of a brightness temperature for the jets, directed almost at the observer.}
{The accurate account for the jet transversal structure allow us to express the magnetic flux through the observed
values and an unknown rotation rate of magnetic surfaces. If we assume the sources are in a magnetically arrested disk state,
the lower limit for the rotation rate can be obtained. On the other hand, the flux estimate may be tested against the
total jet power predicted by the electromagnetic energy extraction model. The resultant expression for power depends logarithmically weakly
on an unknown rotation rate. We show that the total jet power estimated through the
magnetic flux is in good agreement with the observed power. We also obtain the extremely slow rotation rates, which may be an indication
that the majority of the sources considered are not in the magnetically arrested disk state.} 

\tiny
 \keyFont{ \section{Keywords:} active galaxies, jets, BL Lacertae objects, non-thermal radiation, magnetic flux} 
\end{abstract}

\section{Introduction}

One of the key issues of theoretical modeling of relativistic jets is determining the magnetic field magnitude. There are
several theoretical ways to estimate the latter. The {Eddington magnetic field \citep{Beskin-2010}} sets up an upper limit on the magnetic field magnitude 
in the vicinity of a black hole, since it is set by equipartition of magnetic field density and the total energy density in the accreting plasma that is needed to support the Eddington luminosity. The state of magnetically arrested disk (MAD) \citep{NIA-03, TchNMcK-11, McKTchB-12} sets up the limiting magnetic field that can be accreted onto
a black hole basing on an assumption that in such a state the pressure of previously accreted magnetic field can affect the dynamical 
process of accretion itself. 

There are observational means to evaluate the magnetic field in a jet. Blazar spectrum is successfully modeled by the synchrotron self-Compton model. The lower part of the spectrum is dominated by synchrotron radiation of relativistic plasma in a jet magnetic field. Thus, the high-resolution radio interferometry observations
provide us with data for unveiling the physical conditions at the very jet origin --- in so called radio core.
The way to estimate the magnetic field amplitude through the observations is measurements of core shift effect together with
several theoretical assumptions on the radiating volume properties \citep{Lobanov-98}. However, the measurement of radio flux itself, or, equally,
the brightness temperature, can provide us with the instrument to probe the magnetic field magnitude \citep{Zdz-15, Nokhrina-17}.

{The magnetic field estimates is an important parameter that allows to test the theoretical models against observations. \citet{Z-14} used the
magnetic field measurements to calculate the flux and to show that the total flux is in accordance with the magnetically arrested state of the sources.}
In this paper we make use of {the brightness temperature and core shift measurements coupled with the transversal jet model to express the magnetic field magnitude and the magnetic flux contained in a blazar jet through the rotation rate of magnetic surfaces}. Our aim is to compare the magnetic flux in a jet against theoretically limited flux by MAD state and to estimate the rotation rate. {We also test the obtained magnetic flux against the observed jet power. If the jet power $P_{\Psi}$ is fully determined by the electromagnetic energy extraction mechanism, \nt{so we denote it with the subscript $\Psi$}, than it can be expressed as \citep{Beskin-2010}
\begin{equation}
P_{\Psi}=\left(\frac{\Psi a}{\pi r_{\rm g}}\right)^2c,\label{Ptot}
\end{equation} 
where $\Psi$ is the total magnetic flux, $r_{\rm g}$ is a gravitational radius, $c$ is a speed of light, and the rotation rate $a=r_{\rm g}/R_{\rm L}$
is a ratio of a gravitational radius to the light cylinder radius $R_{\rm L}$.}

Although there are estimates for magnetic field amplitude \citep{Lobanov-98, Hirotani-05, OSG-09, 
NBKZ-15, Zdz-15, Nokhrina-17}, it cannot be explicitly used for flux calculation. Indeed, the magneto hydrodynamical theoretical and numerical modeling 
(see e.g. \citet{Lyubarsky-09, BTch-16, NBKZ-15}) show that the toroidal magnetic field is greater than the poloidal one outside the light cylinder radius. Thus, measurements provide us with the magnitude of the toroidal magnetic field, while the poloidal one is needed to estimate the total magnetic flux \citep{Z-14}. 

\section{Magnetic flux in a jet}

{The observed flux, or observed brightness temperature, can be used to estimate the magnetic field in the radiating domain, and, thus,
the magnetic flux. The blazar spectrum in radio band is accurately modeled within the self-absorbed synchrotron model (see e.g. \citet{Abdo-11}). The simplest
model for the source is a uniform sphere with chaotic magnetic field $B$ and relativistic electrons with the power-law energy distribution described
by the amplitude particle number density $k_{\rm e}$ and spectral index $p$: 
\begin{equation}
dn(\gamma)=k_{\rm e}\gamma^{-p}d\gamma,\label{enerdistr}
\end{equation} 
$\gamma\in[\gamma_{\rm min},\,\gamma_{\rm max}]$ \citep{Gould-79}, where $\gamma$ is a Lorentz factor of relativistic plasma.
The observed spectral flux $S_{\nu}$ at the frequency $\nu$ for the optically thick uniform self-absorbed source of radius $R$
at the distance $d$ can be written using the 
spectral photon emission rate $\rho_{\nu}$ and effective absorption coefficient $\ae_{\nu}$ as:
\begin{equation}
S_{\nu}=\pi\hbar\nu\frac{\rho_{\nu}}{\ae_{\nu}}\frac{R^2}{d^2}u(2R\ae_{\nu}),\label{S2}
\end{equation}
and the function of the optical depth $u(2R\ae_{\nu})$ is defined by \citet{Gould-79}. The emission and absorption coefficients for the self-absorbed 
synchrotron model} $\rho_{\nu}$ and $\ae_{\nu}$ are the functions of the magnetic field $B$ and of particle number density amplitude $k_{\rm e}$.
These coefficients, written in a jet frame, i.e. in a frame where the electric field vanishes, are:
\begin{equation}
\rho_{\nu}=4\pi\left(\frac{3}{2}\right)^{(p-1)/2}a(p)\alpha k_{\rm e}\left(\frac{\nu_{\rm B}}{\nu}\right)^{(p+1)/2},\label{rho1} 
\end{equation}
\begin{equation}
\ae_{\nu}=c(p)r_0^2 k_{\rm e}\left(\frac{\nu_0}{\nu}\right)\left(\frac{\nu_{\rm B}}{\nu}\right)^{(p+2)/2}.\label{ae1}
\end{equation}
Here $\nu_{\rm B}=eB/mc$ is a gyrofrequency in the jet frame, $\hbar$ is the Planck constant, $\alpha=e^2/\hbar c$ is 
the fine structure constant, and the functions $a(p)$ and $c(p)$ of the electron
distribution spectral index $p$ are defined in \citep{Gould-79}. {The spectral flux depends on the magnetic field amplitude,
while the particle number density amplitude defines the maximum of a function $u$ and, thus, the position of the observed
radio core --- the domain where the optical depth $\tau$ is equal to unity. So, the spectral flux measurement provide us with an instrument
to evaluate the magnetic field in a source.}
The spectral flux may be expressed through the brightness temperature $T_{\rm b}$ as
\begin{equation}
S_{\nu}=\frac{2\pi\nu^2\theta^2}{c^2}k_{\rm B}T_{\rm b},\label{S1}
\end{equation} 
where $\theta$ is the angular size of a radiating domain. Thus one can express the magnetic field amplitude in a source 
having measured brightness temperature. The method was first applied by \citep{Zdz-15} to check the magnetic field 
amplitude in AGN radio cores independently
of the equipartition assumption. Equating the right-hand sides of equations (\ref{S1}) and (\ref{S2}), and expressing the jet frame values through the
nucleus frame values, we obtain for $p=2$ the magnetic field (\citet{Zdz-15}, see also details in \citet{Nokhrina-17})
\begin{equation}
\left(\frac{B_{\rm uni}}{\rm G}\right)=7.4\cdot 10^{-4}\frac{\Gamma\delta}{1+z}\left(\frac{\nu_{\rm obs}}{\rm GHz}\right)\left(\frac{T_{\rm b,\,obs}}{10^{12}{\rm K}}\right)^{-2}.\label{Eq0}
\end{equation} 
Here $\Gamma$ is a flow bulk Lorentz factor, $z$ is a source redshift, and $\delta$ is a Doppler factor.

{On the other hand, the measurements of core shift effect \citep{Lobanov-98, Hirotani-05, OSG-09, Z-14} provides the
following expression for the magnetic field amplitude $B_{\rm cs}$ at $1$~pc distance from the central source:
\begin{equation}
\left(\frac{B_{\rm cs}}{\rm G}\right)=0.17\left(\frac{\eta_{\rm cs}}{\rm mas\;GHz}\right)^{0.75}\left(\frac{D_{\rm L}}{\rm Gpc}\right)^{0.75}
\frac{\Gamma}{\chi^{0.25}(1+z)^{0.75}\sin^{0.5}\varphi\delta^{0.5}},\label{Bcs}
\end{equation}
here $\eta_{\rm cs}$ is a coefficient determining the slope of the apparent core position dependence on the inverse observation frequency:
\begin{equation}
\left(\frac{\Delta r}{\rm mas}\right)=\left(\frac{\eta_{\rm cs}}{\rm mas\;GHz}\right)\left(\frac{\nu_{\rm obs}}{\rm GHz}\right)^{-1}.
\end{equation}
$\chi$ is the jet opening angle that may obtained having the measured in \citep{Push-09} apparent opening angle $\chi_{\rm app}$ as
$\chi=\chi_{\rm app}\sin\varphi/2$,  $D_{\rm L}$ is a luminosity distance, the bulk plasma motion Lorentz factor is $\Gamma$, $\delta$ is the Doppler factor, and the observation angle $\varphi=\Gamma^{-1}$ \citep{Cohen-07}. The expression (\ref{Bcs}) has been obtained
under the same assumption of synchrotron self-absorbed source, but the method utilizes the core shift effect --- the shift of the observed
radio core on different frequencies. This shift is due to the fact that the surface of optical depth $\tau=1$ is situated at different
distance from the central source for each frequency.
We must stress that the relation (\ref{Eq0}) uses only the synchrotron self-absorbed source model, but the position of the radio core in the model
is not known. If we use the core shift effect, we may also obtain the core position \citep{Nokhrina-17}:
\begin{equation}
\left(\frac{r_{\rm core}}{\rm pc}\right)=\frac{4.85}{\sin\varphi(1+z)^2}\left(\frac{\nu_{\rm obs}}{\rm GHz}\right)^{-1}\left(\frac{\eta_{\rm cs}}{\rm mas\;GHz}\right)
\left(\frac{D_{\rm L}}{\rm Gpc}\right).\label{rcore}
\end{equation}
The relation for $B_{\rm cs}$ (\ref{Bcs}) has been obtained with more assumptions: equipartition between magnetic field energy and plasma energy assumption, 
and the Blanford-K\"onigl model \citep{BK-79} $B(r)=B_1(r_1/r)$, $n(r)=n_1(r_1/r)^2$, where $n_1$ and $B_1$
are particle number density and magnetic field magnitude at distance $r$ along the jet equal to $r_1=1$~pc. 
Relation (\ref{Bcs}) provides the magnetic field amplitude together with its position along the jet.}

Both expressions (\ref{Eq0}) and (\ref{Bcs}) are based on the model of uniform radiating sphere \citep{Gould-79}. In particular, 
such a model does not allow us to estimate the total magnetic flux, contained in a jet --- one of the important
values, defining the total jet power, and the value {that} could be restricted by the magnetically arrested disk model.
{Indeed, as the toroidal magnetic field $B_{\varphi}$ dominates the major part of a jet, it is the toroidal component of a field we imply by
the spectral flux measurement. However, it is the poloidal component $B_{\rm P}$ that carries the magnetic flux.} 
However, the transversal modeling of field profiles allow us in a simple case of blazar jets, i.e. jets pointing almost
directly at us, to calculate the flux from non-homogeneous {cylindrical self-absorbed synchrotron} source, and {to} 
correlate the measured magnetic field amplitude
with the poloidal field in a jet {that} defines the total flux. 
Indeed, it has been shown by \citep{Lyubarsky-09}, that the relation
\begin{equation}
B_{\rm P}=B_{\varphi}\frac{R_{\rm L}}{r_{\perp}}
\end{equation} 
holds outside the light cylinder $R_{\rm L}=c/\Omega_{\rm F}$.
Further we model the transversal magnetic field \nt{and particle number density} profiles as follows.
Inside the light cylinder the poloidal magnetic field remains almost constant \citep{BN-09}, while $B_{\varphi}=B_0 r_{\perp}/R_{\rm L}$.
Both numerical and semi-analytical modeling \citep{NBKZ-15, BTch-16} show, that outside the light cylinder the power-law is a good approximation for 
magnetic field \nt{and particle number density} profiles across the jet for small opening angles. We set 
\begin{equation}
B_{\rm P}=\left\{
\begin{array}{rl}
\displaystyle B_0, & r_{\perp}\le R_{\rm L},\\ \ \\
\displaystyle B_0\left(\frac{R_{\rm L}}{r_{\perp}}\right)^{2}, & R_{\rm L}<r_{\perp}\le R_{\rm j},
\end{array}
\right.\label{Bp}
\end{equation}
\begin{equation}
B_{\varphi}=\left\{
\begin{array}{rl}
\displaystyle B_0\frac{r_{\perp}}{R_{\rm L}}, & r_{\perp}\le R_{\rm L},\\ \ \\
\displaystyle B_0\left(\frac{R_{\rm L}}{r_{\perp}}\right), & R_{\rm L}<r_{\perp}\le R_{\rm j}.
\end{array}
\right.\label{Bphi}
\end{equation}
\begin{equation}
n=\left\{
\begin{array}{rl}
\displaystyle n_0, & r_{\perp}\le R_{\rm L},\\ \ \\
\displaystyle n_0\left(\frac{R_{\rm L}}{r_{\perp}}\right)^{2}, & R_{\rm L}<r_{\perp}\le R_{\rm j},
\end{array}
\right.\label{n}
\end{equation}
\nt{where $B_0$ and $n_0$ are the magnetic field and particle number density amplitudes, i.e. magnitudes at the light cylinder $R_{\rm L}$}.

\nt{For the simplest case of a jet pointing almost at us, when the radiation domain may be treated as a stratified cylinder with the profiles (\ref{Bp})--(\ref{n}), we calculate the spectral flux $S_{\nu}$ (see details in \citep{Nokhrina-17}). Equating the obtained $S_{\nu}$ to the expression (\ref{S1}), 
we obtain the following relation for the amplitude magnetic field $B_0$:   
\begin{equation}
\left(\frac{B_0}{\rm G}\right)=6.4\times 10^{-4}\frac{R_{\rm j}}{R_{\rm L}}\frac{\Gamma\delta}{1+z}
\left(\frac{\nu_{\rm obs}}{\rm GHz}\right)\left(\frac{T_{\rm b,\,obs}}{10^{12}\,{\rm K}}\right)^{-2}.\label{Eq1b}
\end{equation}
Here the fast rotation $\Gamma R_{\rm L}\ll R_{\rm j}$ is assumed.}
While $R_{\rm j}$ can be estimated through observations, the light cylinder radius is usually unknown. 
However, its value can be somewhat restricted by theoretical models predicting that Blandford-Znajek process
works effectively for $R_{\rm L}$ of the order of $2 r_{\rm g}$ \citep{BZ-77, TchMcKN-12}. Still, the value $B_0$ cannot be readily extracted from the observations.

{The total flux $\Psi$ in a jet with given cross-section magnetic filed profile is defined as
\begin{equation}
\Psi=2\pi\int_0^{R_{\rm j}}B_{\rm P}r_{\perp}dr_{\perp}.
\end{equation}
Using the magnetic field profile (\ref{Bp}) we obtain
\begin{equation}
\Psi=\pi B_0 R_{\rm L}^2\left(1+2\ln{\frac{R_{\rm j}}{R_{\rm L}}}\right).\label{Psi_v1}
\end{equation}
Substituting explicitly $B_0 R_{\rm L}$ into (\ref{Psi_v1}) and using the correlation 
$B_0 R_{\rm L}=0.86B_{\rm uni}R_{\rm j}$ following from (\ref{Eq0}) and (\ref{Eq1b}),
we obtain the relation for the magnetic flux:
\begin{equation}
\Psi=2.7 B_{\rm uni,\;cs} R_{\rm j} \frac{r_{\rm g}}{a}\left[1+2\ln{\frac{R_{\rm j}\,a}{r_{\rm g}}}\right]=\frac{\Psi_a}{a}. 
\label{Psib}
\end{equation}
Here one may use $B_{\rm cs}$ or $B_{\rm uni}$, since both values are obtained under the same assumptions on the
geometry and structure of radiating domain.} \nt{The amplitude magnetic flux $\Psi_{\rm a}=a\Psi$.}
The equation (\ref{Psib}) coincides with the expression for the magnetic flux in \citep{Z-14}.
The expression (\ref{Psib}) for the flux depends inversely on a rotation rate $a$, because the dependence on
the physical values in square brackets is logarithmically weak and can be neglected. Taking the fiducial value for $R_{\rm j}/r_{\rm g}\sim 10^3$, 
we take the expression in square brackets being of the order of a few to ten. 


\section{The jet power and the rotation rate}

We apply the obtained expression for the flux (\ref{Psib}) to test it against the following theoretical predictions.
If we assume that most of the sources are in MAD state, we can compare the amplitude flux $\Psi_{\rm a}$ and $\Psi_{\rm MAD}$ and obtain the rotation parameter 
$a=\Psi_{\rm a}/\Psi_{\rm MAD}$.
However, we must bear in mind that the energy losses mechanism \citet{BZ-77} works effectively for relatively high
rotation rates $a>0.5$. 
Thus, with the difference in $\Psi_{\rm MAD}$ and $\Psi_{\rm a}$ is greater, we might think that the source is not in the MAD state.

We also use the obtained flux (\ref{Psib}) to calculate the total jet power given by the equation (\ref{Ptot}). As the expression (\ref{Ptot}) depends
on the product $\Psi a=\Psi_{\rm a}$ (\ref{Psib}) that depends on $a$ logarithmically weakly only through the term $1+2\ln(R_{\rm j}a/r_{\rm g})$, so does the total power. So this result 
is independent on the assumption of the particular value for $a$. That is why such a test may be important for the flux determination.
We do it for the magnetic field estimated by two methods: the brightness temperature measurements and the core shift measurements in order 
to compare the two methods.


{We have found 48 sources meeting the following conditions: i. the observational angle of a jet
must be small enough for the model of head-on jet for $B_{\rm uni}$ can be applied; ii. the source has a measured core shift,
central black hole mass, and the apparent opening angle. We use
the following samples: we take the brightness temperature measured by \citet{Kovalev-05} and the core shifts by \citet{Push-12}.
The apparent velocity $\beta_{\rm app}$ we take from \citep{MOJAVE-09}.
We also use the black hole masses $M_{\rm BH}$ and the accretion luminosities $L_{\rm acc}$ collected by \citet{Z-14}. For the
unknown Lorentz factor we use $\Gamma=\sigma_{\rm M}$, and the Michel's magnetization parameter $\sigma_{\rm M}$ has been evaluated by \citet{NBKZ-15}, \nt{We use for the observed 
opening angle the results from \citet{Push-09}. This is in contrast with \citep{Z-14}, where the causal connectivity across the jet
$\Gamma\chi\sim 1$ is assumed.}
{We obtain the magnetic field magnitudes using the above values and equations (\ref{Eq0}) and (\ref{Bcs}).
On average, the values of $B_{\rm uni}$ is less and more scattered than values of $B_{\rm cs}$, which is in agreement with results obtained in \citep{Zdz-15}.
To calculate the flux using $B_{\rm cs}$ we employ $R_{\rm j}=\chi \times 1\;{\rm pc}$. 
For $B_{\rm uni}$ we use equation (\ref{rcore}), so we define $R_{\rm j}=r_{\rm core}\chi$.}

One of the possible upper limits on the magnetic field amplitude in 
relativistic jets may be imposed by 
MAD model. Magnetically arrested disk is a disk in a state of equilibrium of the accretion rate and
the pressure of magnetic field frozen in previously accreted matter\citep{NIA-03, TchNMcK-11, McKTchB-12}. There is observational support of
AGNs staying in MAD state \citep{Z-14}. Equation  (\ref{Psib}) {relate} the total magnetic 
flux in a jet with the observable jet radius, magnetic field, gravitational radius 
(through the black hole mass estimates), and unknown rotation rate $a$. 
{Setting $\Psi_{\rm MAD}$ as the upper limit on a magnetic flux, 
one can obtain the lower limit for the rotational rate of a central black hole.}
Here we compare the magnetic flux $\Psi$ calculated with expression (\ref{Psib}) for $B_{\rm uni}$, obtained by the
brightness temperature measurements, or $B_{\rm cs}$ obtained by core shift measurements, and the magnetic flux $\Psi_{\rm MAD}$ set by MAD model.  
In order to obtain the magnetic flux predicted by MAD $\Psi_{\rm MAD}$, we use 
the equation \citep{Z-14}
\begin{equation}
\Psi_{\rm MAD}\approx 50\sqrt{\dot{M}r_{\rm g}^2 c},\label{Ppsi}
\end{equation} 
where we use relation between disk luminosity $L_{\rm acc}=\eta\dot{M}c^2$. 

\nt{The results for $\Psi$ calculated for $a=0.5$ and
$\Psi_{\rm MAD}$ are presented in Table~1. We present in the table the values for $\Psi_{\rm MAD}$ \citep{Z-14},
the total magnetic flux obtained using brightness temperature measurements $\Psi_{\rm br}$ and core shift measurements $\Psi_{\rm cs}$.
We see the reasonable agreement between $\Psi_{\rm br}$ and $\Psi_{\rm cs}$, although the former is more scattered.
We see that $a\Psi\ll\Psi_{\rm MAD}$ for almost all the sources
for magnetic filed estimates by both methods.
If we assume that all the sources are in MAD state, the rotational rate of a black hole must be in a range $(0.0001;\;0.1)$ for 36 sources. Only
12 have the rotational rate between $0.1$ and $1$. Thus, assumption of the sources being in the MAD state leads us to a conclusion
that the rotation must be much less than the critical one.}

\nt{Otherwise, we may assume that 36 sources have rotation parameters close to critical $a\in[0.5;\;1]$, but not in a MAD state.
We must stress that the expressions for the magnetic field estimate through the core shift measurement are different here 
from the one used in \citep{Z-14}. In this paper the Eq.~(\ref{Bcs}) uses the assumtions from \citep{NBKZ-15} of the total
outflow magnetization equal to the unity. This condition means that the total Poynting flux in a core region is equal to
the total bulk particle kinetic energy flux, with about $1\%$ \citep{SSA-13} of radiating particles having the relativistic energy distribution (\ref{enerdistr}).
This assumption has been used to estimate maximal possible Lorentz factor of the flow \citep{NBKZ-15}, which correlates
very good with the Lorentz factor estimates basing the observed super-luminal velocities \citep{MOJAVE-09}. In this point our approach differs
from the one used by \citep{Z-14}.}


{In order to check our flux estimates, we test it against the total jet power (\ref{Ptot}).
This result is robust under the model assumptions, since it the total power depends on $a$ very weakly. 
The calculation of the total jet power for the obtained flux is in Table~1. We compare the total power $P_{\Psi}$, calculated \nt{substituting
(\ref{Psib}) into (\ref{Ptot}), with the 
jet power, estimated basing on the correlation of $P_{\rm jet}$ with the luminosities of jet radio band \citet{Cav-10}:
\begin{equation}
\left(\frac{P_{\rm jet}}{10^{43}\;\rm erg\;s^{-1}}\right)=3.5\left(\frac{P_{200-400}}{10^{40}\;\rm erg\;s^{-1}}\right)^{0.64}.\label{Pobs}
\end{equation}
We plot $P_{\Psi}$ against the 
obtained with (\ref{Pobs}) power $P_{\rm jet}$ in Fig.~1. We observe the reasonable correlation 
of $P_{\Psi}$ and $P_{\rm jet}$. The histogram of the ratio of $P_{\Psi}/P_{\rm jet}$ is presented in Fig.~2.
We see that the ratio has a well determined peak around a few. Although it gives the systematic excess
of $P_{\Psi}$ over $P_{\rm jet}$, we state that $P_{\Psi}$ is in accordance with $P_{\rm jet}$ bearing in mind uncertainties
in the determination of all the values including $P_{\rm tot}$. The systematic excess may be attributed to the probable overestimating the magnetic field $B_{\rm cs}$,
the hints of which we see in discrepancy between $B_{\rm cs}$ and $B_{\rm br}$, the latter being lower.}    


We have also tested the jet power obtained with the flux determined by $B_{\rm uni}$ (see Fig.~1).
The second method provides systematically lower powers and more scattering. This is in agreement with the result by \citep{Zdz-15} who have found the scattering
in magnetic field amplitude calculated with no equipartition assumption while still have the majority of sources having the equipartition magnetic field.}

\section{Summary}

We have discussed estimates for the magnetic flux using the jet core magnetic field obtained through
{the brightness temperature measurements} and by the core shift effect. 
Usually, any estimates of the magnetic field in a radiating
domain of relativistic jet cannot be readily put into the expression for the magnetic flux. This is because
theoretical modeling show, that the toroidal magnetic field dominates the poloidal magnetic field outside the light
cylinder. Thus, the field we measure using synchrotron self-Compton model of radiation, must be toroidal,
while the magnetic flux {is} determined by the poloidal one. Consideration of transversal field configuration is needed
to estimate accurately the magnetic flux in a jet using the available evaluation of magnetic field magnitude
through observations --- either using the core-shift effect, or spectral flux measurements. In this work
has been considered the simplest case --- the transversal structure of jets  observed with very small
viewing angle.  

We test the method of estimating the flux against the limiting flux determined by the magnetically arrested disk model.
For 36 of 48 sources the obtained flux is much less than the MAD flux. This suggest either the extremely slow rotation 
rate $a\in(0.0001;\;0.1)$ or that the sources are not in MAD state. For 12 sources both fluxes coincide for $a\in(0.1;\;1)$ ---
the fast rotation that is needed for the efficient energy extraction from a black hole.

We also test the flux estimate against the total jet power determined by the electromagnetic mechanism of energy 
extraction. \nt{This result does not depend on the particular value for $a$, as the expression (\ref{Ptot}) depends on the product
of $\Psi$ and $a$ $\Psi_{\rm a}$ that can be estimated directly. In this case we see a good agreement between the total power determined by the flux and the total power 
obtained from the observations, with the distribution of powers ration being well peaked around a few.}


\section*{Author Contributions}

The results presented in the paper has been obtained by EN.

\section*{Funding}
This work was supported by Russian Science Foundation, grant 16-12-10051.

\section*{Acknowledgments}
The author thanks the anonymous referees for
the suggestions which helped to improve the paper.
{This research has made use of data from the MOJAVE database that is maintained by the MOJAVE team \citep{MOJAVE-09}.}


\begin{figure}
\includegraphics[scale=0.5]{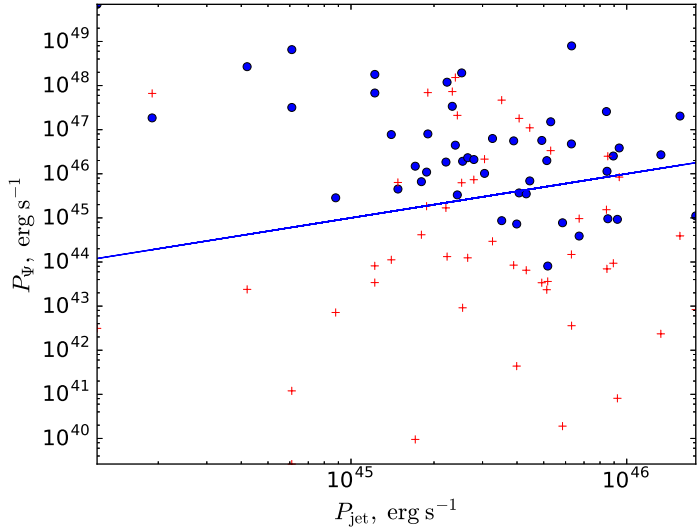} 
\caption{The jet power $P_{\Psi}$, calculated using magnetic flux, against the total kinetic jet power. The straight line
is a theoretical prediction. The blue circle stand for the total flux obtained using the core shift effect, the red crosses are
for the total flux obtained using the brightness temperature. The sources with the flux approximately equal to the MAD flux are in the upper
left corner.} 
\label{fig_1} 
\end{figure}

\begin{figure}
\includegraphics[scale=0.5]{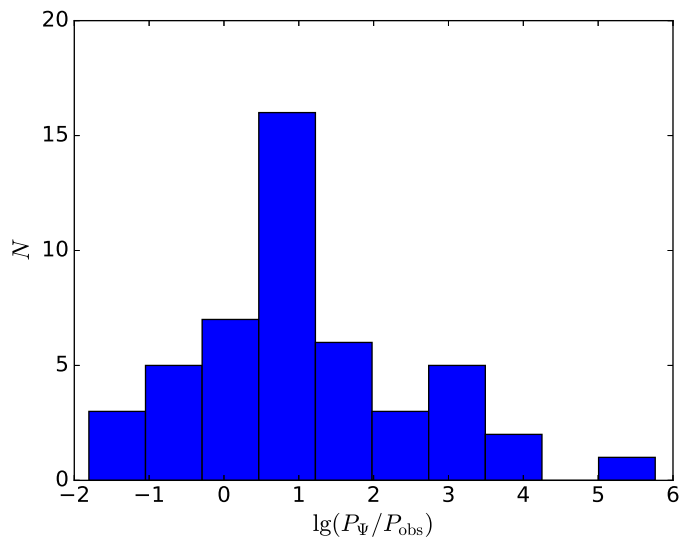} 
\caption{The histogram showing the number of sources with the ration of calculated power $P_{\Psi}$ to the total jet power $P_{\rm jet}$, the ratio
is in the log-scale. We see the systematic excess of power estimated through the flux against the jet power by a factor of few.} 
\label{fig_1}
\end{figure}

\begin{table*}
 \begin{minipage}{150mm}
  \caption{Jet important observed and derived parameters.\label{table}}
  \begin{tabular}{ccccccc}
  \hline
  Source & z & $\Psi_{\rm MAD}$   & $\Psi_{\rm br}$ & $\Psi_{\rm cs}$ & $P_{\Psi}$            & $P_{\rm jet}$         \\
         &   & ${\rm G\; cm^{2}}$ & ${\rm G\; cm^{2}}$  & ${\rm G\; cm^{2}}$  & $[{\rm erg\;s^{-1}}]$ & $[{\rm erg\;s^{-1}}]$ \\
     (1) & (2) & (3)              & (4)                 & (5)                 & (6)                   & (7)                   \\
\hline
0133$+$476 & $0.859$ & $5.51\times 10^{33}$ & $1.17\times 10^{31}$ & $5.34\times 10^{32}$ & $1.92\times 10^{46}$ & $2.54\times 10^{45}$ \\
0212$+$735 & $2.367$ & $5.77\times 10^{35}$ & $5.97\times 10^{32}$ & $8.93\times 10^{32}$ & $8.10\times 10^{43}$ & $5.17\times 10^{45}$ \\
0234$+$285 & $1.206$ & $5.71\times 10^{34}$ & $1.24\times 10^{34}$ & $5.31\times 10^{32}$ & $8.65\times 10^{44}$ & $3.52\times 10^{45}$ \\
0333$+$321 & $1.259$ & $9.36\times 10^{34}$ & $6.00\times 10^{32}$ & $3.81\times 10^{32}$ & $3.88\times 10^{44}$ & $6.72\times 10^{45}$ \\
0336$-$019 & $0.852$ & $1.55\times 10^{34}$ & $1.45\times 10^{32}$ & $2.12\times 10^{33}$ & $6.31\times 10^{46}$ & $3.26\times 10^{45}$ \\
0403$-$132 & $0.571$ & $3.00\times 10^{34}$ & $4.34\times 10^{33}$ & $1.09\times 10^{33}$ & $6.89\times 10^{45}$ & $4.45\times 10^{45}$ \\
0528$+$134 & $2.070$ & $6.05\times 10^{34}$ & $1.61\times 10^{30}$ & $3.24\times 10^{32}$ & $7.75\times 10^{44}$ & $5.85\times 10^{45}$ \\
0605$-$085 & $0.870$ & $1.68\times 10^{34}$ & $9.94\times 10^{33}$ & $1.70\times 10^{33}$ & $4.45\times 10^{46}$ & $2.39\times 10^{45}$ \\
0736$+$017 & $0.189$ & $6.94\times 10^{32}$ & $3.86\times 10^{30}$ & $1.29\times 10^{33}$ & $2.68\times 10^{48}$ & $4.20\times 10^{44}$ \\
0738$+$313 & $0.631$ & $1.48\times 10^{35}$ & $3.22\times 10^{33}$ & $2.71\times 10^{33}$ & $4.51\times 10^{45}$ & $1.48\times 10^{45}$ \\ 
0748$+$126 & $0.889$ & $4.33\times 10^{34}$ & $1.39\times 10^{32}$ & $1.90\times 10^{33}$ & $2.31\times 10^{46}$ & $2.65\times 10^{45}$ \\
0827$+$243 & $0.943$ & $1.81\times 10^{34}$ & $1.72\times 10^{32}$ & $6.87\times 10^{32}$ & $6.62\times 10^{45}$ & $1.80\times 10^{45}$ \\
0836$+$710 & $2.218$ & $1.78\times 10^{35}$ & $7.19\times 10^{31}$ & $8.30\times 10^{32}$ & $1.11\times 10^{45}$ & $1.78\times 10^{46}$ \\
0906$+$015 & $1.026$ & $9.81\times 10^{33}$ & $5.66\times 10^{32}$ & $3.90\times 10^{32}$ & $1.02\times 10^{46}$ & $3.05\times 10^{45}$ \\
0917$+$624 & $1.453$ & $2.25\times 10^{34}$ & $3.93\times 10^{33}$ & $5.62\times 10^{32}$ & $3.68\times 10^{45}$ & $4.07\times 10^{45}$ \\
0945$+$408 & $1.249$ & $2.27\times 10^{34}$ & $1.29\times 10^{32}$ & $2.32\times 10^{33}$ & $4.75\times 10^{46}$ & $6.30\times 10^{45}$ \\
1038$+$064 & $1.265$ & $4.33\times 10^{34}$ & $1.16\times 10^{32}$ & $8.49\times 10^{32}$ & $3.51\times 10^{45}$ & $4.32\times 10^{45}$ \\
1127$-$127 & $1.184$ & $7.44\times 10^{34}$ & $2.10\times 10^{32}$ & $3.45\times 10^{33}$ & $2.53\times 10^{46}$ & $8.94\times 10^{45}$ \\
1156$+$295 & $0.725$ & $6.33\times 10^{33}$ & $3.48\times 10^{31}$ & $8.91\times 10^{32}$ & $5.58\times 10^{46}$ & $3.89\times 10^{45}$ \\
1219$+$285 & $0.103$ & $2.83\times 10^{32}$ & $4.34\times 10^{33}$ & $2.29\times 10^{33}$ & $1.85\times 10^{47}$ & $1.90\times 10^{44}$ \\
1222$+$216 & $0.434$ & $1.50\times 10^{34}$ & $6.71\times 10^{33}$ & $2.28\times 10^{33}$ & $8.01\times 10^{46}$ & $1.90\times 10^{45}$ \\
1253$-$055 & $0.536$ & $2.76\times 10^{33}$ & $3.93\times 10^{30}$ & $5.84\times 10^{33}$ & $7.93\times 10^{48}$ & $6.31\times 10^{45}$ \\
1308$+$326 & $0.997$ & $1.11\times 10^{34}$ & $4.89\times 10^{32}$ & $8.28\times 10^{32}$ & $2.10\times 10^{46}$ & $2.79\times 10^{45}$ \\
1334$-$127 & $0.539$ & $1.28\times 10^{33}$ & $1.01\times 10^{29}$ & $1.27\times 10^{32}$ & $1.49\times 10^{46}$ & $1.71\times 10^{45}$ \\
1458$+$718 & $0.904$ & $5.84\times 10^{34}$ & $2.83\times 10^{31}$ & $3.03\times 10^{33}$ & $2.70\times 10^{46}$ & $1.33\times 10^{46}$ \\
1502$+$106 & $1.839$ & $2.17\times 10^{34}$ & $3.48\times 10^{31}$ & $1.43\times 10^{33}$ & $5.70\times 10^{46}$ & $4.92\times 10^{45}$ \\
1510$-$089 & $0.360$ & $2.30\times 10^{33}$ & $1.01\times 10^{31}$ & $1.42\times 10^{33}$ & $6.79\times 10^{47}$ & $1.22\times 10^{45}$ \\
1546$+$027 & $0.414$ & $5.84\times 10^{33}$ & $1.11\times 10^{30}$ & $8.20\times 10^{33}$ & $6.53\times 10^{48}$ & $6.10\times 10^{44}$ \\
1606$+$106 & $1.232$ & $2.44\times 10^{34}$ & $1.86\times 10^{33}$ & $3.95\times 10^{33}$ & $1.51\times 10^{47}$ & $5.30\times 10^{45}$ \\
1611$+$343 & $1.400$ & $5.21\times 10^{34}$ & $6.59\times 10^{32}$ & $8.55\times 10^{33}$ & $2.57\times 10^{47}$ & $8.44\times 10^{45}$ \\
1633$+$382 & $1.813$ & $4.53\times 10^{34}$ & $1.20\times 10^{32}$ & $1.53\times 10^{33}$ & $1.14\times 10^{46}$ & $8.48\times 10^{45}$ \\
1637$+$574 & $0.751$ & $5.51\times 10^{34}$ & $7.76\times 10^{32}$ & $1.89\times 10^{33}$ & $1.09\times 10^{46}$ & $1.88\times 10^{45}$ \\
1641$+$399 & $0.593$ & $5.64\times 10^{34}$ & $9.82\times 10^{31}$ & $2.86\times 10^{33}$ & $1.99\times 10^{46}$ & $5.13\times 10^{45}$ \\
1655$+$077 & $0.621$ & $1.65\times 10^{32}$ & $1.77\times 10^{32}$ & $1.21\times 10^{32}$ & $3.39\times 10^{47}$ & $2.33\times 10^{45}$ \\
1749$+$096 & $0.322$ & $7.70\times 10^{33}$ & $3.28\times 10^{29}$ & $3.62\times 10^{33}$ & $3.19\times 10^{47}$ & $6.10\times 10^{44}$ \\
1803$+$784 & $0.680$ & $1.11\times 10^{33}$ & $1.04\times 10^{31}$ & $9.88\times 10^{32}$ & $1.19\times 10^{48}$ & $2.23\times 10^{45}$ \\
1823$+$568 & $0.664$ & $7.61\times 10^{32}$ & $7.49\times 10^{31}$ & $1.32\times 10^{33}$ & $1.93\times 10^{48}$ & $2.52\times 10^{45}$ \\
1828$+$487 & $0.692$ & $5.21\times 10^{33}$ & $5.52\times 10^{31}$ & $1.26\times 10^{33}$ & $2.04\times 10^{47}$ & $1.56\times 10^{46}$ \\
1849$+$670 & $0.657$ & $1.14\times 10^{34}$ & $6.34\times 10^{31}$ & $9.40\times 10^{33}$ & $1.79\times 10^{48}$ & $1.22\times 10^{45}$ \\
1928$+$738 & $0.302$ & $5.27\times 10^{33}$ & $2.58\times 10^{31}$ & $6.78\times 10^{32}$ & $7.75\times 10^{46}$ & $1.40\times 10^{45}$ \\
2121$+$053 & $1.941$ & $3.10\times 10^{34}$ & $4.34\times 10^{30}$ & $1.77\times 10^{32}$ & $7.27\times 10^{44}$ & $3.99\times 10^{45}$ \\
2155$-$152 & $0.672$ & $3.64\times 10^{32}$ & $1.94\times 10^{32}$ & $2.44\times 10^{31}$ & $3.32\times 10^{45}$ & $2.43\times 10^{45}$ \\
\hline
\end{tabular}
\end{minipage}
\end{table*}

\begin{table*}
 \centering
 \begin{minipage}{150mm}
  \begin{tabular}{ccccccc}
  \hline
2200$+$420 & $0.069$ & $9.05\times 10^{32}$ & $3.27\times 10^{30}$ & $1.54\times 10^{34}$ & $6.95\times 10^{49}$ & $1.20\times 10^{44}$ \\
2201$+$315 & $0.295$ & $2.61\times 10^{34}$ & $2.54\times 10^{31}$ & $5.06\times 10^{32}$ & $2.85\times 10^{45}$ & $8.80\times 10^{44}$ \\
2230$+$114 & $1.037$ & $2.27\times 10^{34}$ & $2.64\times 10^{30}$ & $2.82\times 10^{32}$ & $9.27\times 10^{44}$ & $9.25\times 10^{45}$ \\
2251$+$158 & $0.859$ & $1.81\times 10^{34}$ & $4.89\times 10^{32}$ & $1.05\times 10^{33}$ & $3.85\times 10^{46}$ & $9.39\times 10^{45}$ \\
2345$-$167 & $0.576$ & $3.36\times 10^{33}$ & $1.32\times 10^{32}$ & $4.36\times 10^{32}$ & $1.84\times 10^{46}$ & $2.21\times 10^{45}$ \\
2351$+$456 & $1.986$ & $5.84\times 10^{34}$ & $3.36\times 10^{33}$ & $6.56\times 10^{32}$ & $9.56\times 10^{44}$ & $8.54\times 10^{45}$ \\
\hline
\end{tabular}
\textit{Notes.} 
Columns are as follows: (1) source name (B1950); (2) redshift $z$ as collected by \citep{Listeretal-13}; (3) the MAD magnetic flux obtained using Equation (\ref{Ppsi}); (4) derived total magnetic flux using the brightness temperature measurements; (5) derived total magnetic flux using the core shift measurements; \nt{(6) The jet power estimate using the
magnetic field $B_{\rm cs}$}; (7) the total jet power estimated
using the correlation between the jet power and radio flux \citep{Cav-10}, collected from \citep{NBKZ-15}.
\end{minipage}
\end{table*}

\bibliographystyle{frontiersinSCNS_ENG_HUMS} 
\bibliography{Nokhrina_bib}







\end{document}